\documentclass[twocolumn,prb,aps,showpacs,superscriptaddress]{revtex4}
\usepackage{graphicx}

\begin{document}

\title{Flux-flow resistivity anisotropy in the instability regime
in the \textit{a-b} plane of epitaxial $YBa_2Cu_3O_{7-\delta}$
thin films}

\author{B. Kalisky}
\affiliation{Department of physics, Institute of
Superconductivity, Bar-Ilan University, Ramat-Gan 52900, Israel}
\affiliation{Department of condensed matter physics, Weizmann
Institute of Science, Rehovot, 76100, Israel}

\author{P. Aronov}
\affiliation{Department of Physics, Technion - Israel Institute of
Technology, Haifa 32000, Israel}

\author{G. Koren}
\email{gkoren@physics.technion.ac.il}\affiliation{Department of
Physics, Technion - Israel Institute of Technology, Haifa 32000,
Israel}

\author{A. Shaulov}
\affiliation{Department of physics, Institute of
Superconductivity, Bar-Ilan University, Ramat-Gan 52900, Israel}

\author{Y. Yeshurun}
\affiliation{Department of physics, Institute of
Superconductivity, Bar-Ilan University, Ramat-Gan 52900, Israel}

\author{R. P. Huebener}
\affiliation{Institute of Physics, University of Tuebingen,
Morgenstelle 14, D-72076 Tuebingen, Germany}

\begin{abstract}

Measurements of the nonlinear flux-flow resistivity $\rho$ and the
critical vortex velocity $\rm v^*_\varphi$ at high voltage bias
close to the instability regime predicted by Larkin and
Ovchinnikov \cite{LO} are reported along the node and antinode
directions of the d-wave order parameter in the \textit{a-b} plane
of epitaxial $YBa_2Cu_3O_{7-\delta}$ films. In this pinning-free
regime, $\rho$ and $\rm v^*_\varphi$ are found to be anisotropic
with values in the node direction larger on average by 10\% than
in the antinode direction. The anisotropy of $\rho$ is almost
independent of temperature and field. We attribute the observed
results to the anisotropic quasiparticle distribution on the Fermi
surface of $YBa_2Cu_3O_{7-\delta}$.

\end{abstract}


\maketitle

\indent Flux flow due to the Lorentz force in a current carrying
type-II superconductor in the mixed state has been studied
intensively for many years \cite{Huebener,Tinkham}. In view of the
$d_{x^2-y^2}$-wave symmetry of the pair wavefunction in the
hole-doped cuprates \cite{Tsuei}, measurements of the flux-flow
resistance (FFR) as a function of the crystallographic direction
within the $CuO_2$ planes become highly interesting. In
\textit{c-axis} oriented epitaxial $YBa_2Cu_3O_{7-\delta}$ (YBCO)
films under an applied magnetic field parallel to the c-axis, one
could envision an anisotropic behavior resulting in some
difference between the flux motion in the node and antinode
directions. However, up to now no such anisotropy has been
reported, nor has such an effect been analyzed theoretically.
Recently, a search study for flux flow resistivity anisotropy in
the low bias pinning regime was carried out by some of us in
epitaxial thin films of YBCO \cite{Koren}. To within the
experimental error in that study, no anisotropy was found between
the transport properties of microbridges patterned along the node
and antinode directions. Since at low bias the transport results
are determined by the pinning properties of the films, it was
concluded that the pinning properties in YBCO are isotropic. At
high bias however, close to the Larkin Ovchinikkov (LO)
instability regime \cite{LO}, pinning is negligible and flux flow
resistivity measurements could reveal an intrinsic anisotropy
between the node and antinode directions. In this study we present
the first experimental results indicating such anisotropy in the
\textit{a-b} plane of YBCO films, where the nonlinear FFR along
the node direction is enhanced on average by about 10\% as
compared to that along the antinode direction.\\
\indent Our measurements were performed on the same two wafers
used previously in the low bias measurements \cite{Koren}. We used
two high quality epitaxial, \textit{c-axis} oriented YBCO films of
0.12$\,\mu m$ thickness prepared under identical conditions by
laser ablation deposition on (100) $SrTiO_3$ (STO) wafers of
$10\times 10\, mm^2$ area. The films were patterned with a
photolithographic mask into 10 equally spaced $0.12\times 12\times
100\,\mu m^3$ bridges along the line dividing the wafer into two
halves by deep UV photolithography and Ar ion milling. The
dimensions of the resulting bridges were measured with optical and
atomic force microscopes and found to conform with the mask design
to within $\pm 0.4\%$. Successive microbridges were oriented at
alternating angles of $0^\circ$ and $45^\circ$ to the edge of the
wafers as shown schematically in the upper panel of Fig. 1, so
that the transport current would flow either along the node or the
antinode directions of the order parameter.
\begin{figure}
\includegraphics[width=8.2cm]{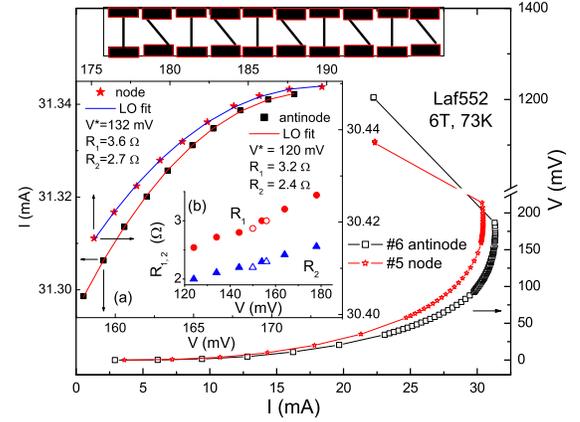}
\caption{(Color online) A schematic drawing of the bridges on the
wafers, and a node and antinode bridges voltage V versus current I
at B = 6 T and T = 73 K. Inset (a) shows the corresponding I
versus V just below the maximum of I versus V with fits using Eq.
(1), inset (b) shows the dynamic nature of $R_1$ and $R_2$, see
text. }\label{fig1}
\end{figure}
\noindent The alternating direction of adjacent bridges is
important in order to minimize systematic differences due to
possible large-scale inhomogeneities in the films. On one film,
where the orientation of the STO substrate was with the edge of
the wafer parallel to the (010) crystalline direction, five odd
number bridges were along the antinode direction, and five even
number bridges along the node direction. In the other film where
the side of the wafer was parallel to the (110) orientation, the
role of the antinode and node bridges was reversed due to the
epitaxial growth of the film.  Studying these two types of wafers
was done in order to check if our ion milling process, done at an
incident angle of 45$^\circ$ to the wafers, is affecting the
properties of the bridges. Any observed difference in the
transport properties of the two wafers would imply that the effect
is not intrinsic, and results from the patterning process. Low
resistance gold contacts were used, and the wafers were annealed
for optimal oxygen doping of the films. The resistance
measurements were done by the standard four-point dc technique in
Helium gas environment with and without a magnetic field of up to
8 T normal to the wafers.\\
\indent The low bias normal-state resistivity  $\rho$ under
zero-magnetic-field was measured versus temperature for each
bridge on the two wafers. Averaged values of the innermost bridges
are given in Figs. 1 and 2 of Ref. 5. The onset of the
superconductive transition at B = 0 was identical for the two
kinds of bridges on the two wafers, with $T_c^{onset}$ = 94 K.
This value of $T_c^{onset}$ has also been reported by Gross
\textit{et al.} \cite{Gross} for the same kind of films. The
resistivity in the normal state at 95 K was $\rho$(95 K) = 168
$\mu\Omega cm$, being identical for both wafers and both type of
bridges. The observation that  $\rho$(node) = $\rho$(antinode)
results from the four-fold symmetry and the heavy twinning of our
films.\\
\indent At lower temperatures, the FFR in the pinning regime of
YBCO was recently found to be isotropic \cite{Koren}. To determine
the pinning-independent FFR we measured the current-voltage
characteristic near the LO instability caused by the
nonequilibrium distribution of the quasiparticles at moderate
electric fields \cite{LO}. This instability has been studied in
detail both in conventional superconductors and in the cuprates
\cite{Huebener}. Following procedures of earlier experiments
\cite{Dottinger}, for each temperature and magnetic field, the
voltage applied to the sample was ramped up until a resistive jump
due to the LO instability was reached. A typical such jump is seen
in the I-V curves of Fig. 1. Above this jump a thermal runaway is
observed (not shown here), where the bridges reach the normal
state with a typical resistivity of 200 $\mu\Omega cm$ which
corresponds to about $\rho(105\,K)$. We normally did not expose
our bridges to this regime to avoid possible thermal damage. Eq.
(53) of the LO theory \cite{LO} yields
\begin{equation}
I =\frac {V}{R_1}\left\{1+c\, \sqrt{1-\frac{T}{T_c}}\right\}-
\frac{V}{R_{2}}\left(\frac{V}{V^*}\right)^2\left\{ 1+
\left(\frac{V}{V^*}\right)^2\right\}^{-1}
\end{equation}
\noindent where $c$ is a number of order unity, $V^*$ is a
critical voltage, and $R_1$ and $R_{2}$ are two FFRs which were
taken equal in the original theory. We stress that Eq. (1) is
valid only in the flux flow regime at high bias near the maximum
of I versus V, and \textit{not} in the pinning regime at low bias.
In the low bias regime where $I$ is linear in $V$, the FFR was
found to be isotropic \cite{Koren}, as is expected for a heavily
twinned YBCO crystal with a four-fold symmetry if the pinning is
also isotropic. By fitting our I-V data in the instability regime
using Eq. (1) we find that the FFR $R_1/(1+\sqrt{1-T/T_c})$ and
$R_2$ have similar values to within 10\%. Since the Lorentz force
is oriented perpendicular to the current and the applied magnetic
field, and YBCO has a four-fold symmetry, flux flow is along the
node direction in the node bridges, and in the antinode direction
in the antinode bridges. The resistive voltage across the bridge
is $V=|\textbf{v}_\varphi\times\textbf{B}|L$, where
$\textbf{v}_\varphi$ is the flux-flow velocity, \textbf{B} is the
magnetic field and $L$ is the bridge length. The critical voltage
$V^*$ in Eq. (1) is thus associated with a critical vortex
velocity $\textbf{v}^*_\varphi$ by this relation.\\
\indent Joule heating of our samples could affect our results. The
possible temperature rise is $\Delta T=\alpha P$, where \textit{P}
is the dissipated electric power per unit area of the sample, and
$\alpha$ is the thermal boundary resistance \cite{Marshall}. YBCO
films on $SrTiO_3$ substrates have $\rm \alpha \cong 10^{-4} \,
cm^2 K/W$ at 300 K \cite{Marshall}. Assuming an even larger value
of $\rm 10^{-3}\,  cm^2 K/W$, and taking the maximum values of
\textit{P} close to the LO instability, we estimate that $\Delta
T$ remains smaller than a few tenth of a Kelvin. We therefore
conclude that Joule heating of our samples can be neglected,
except perhaps close to $T_c$. Experimentally, different rates of
the current sweeps (14 - 300 mA/s) showed reproducibility of the
I-V data, thus excluding the possibility of magnetic hysteresis
\cite{Beena} and thermal heating effects.\\
\indent Two typical I-V curves for a node and antinode bridges are
shown in Fig. 1 at T = 73 K and B = 6 T. The difference between
the two curves is very clear and typical of most of our bridges.
With increasing current above the critical current regime, the
curve becomes steeper and steeper up to the instability. In the
case of voltage-controlled operation as we have, a negative
differential resistance sets in just before the jump as seen in
Fig. 1. In the pinning-independent instability regime, the
flux-flow process follows closely Eq. (1). Indeed, in inset (a) to
Fig. 1 where ten data points of $I$ versus $V$ just \textit{below}
the maximum of $I$ are plotted for each curve, one can see that
the fits using Eq. (1) are in excellent agreement with the data.
In the fits we assumed that $c=1$, and obtained $R_1$, $R_2$ and
$V^*$. In the following we use the
procedure as in inset (a) to Fig. 1, to present our results.\\
\begin{figure}
\includegraphics[width=8cm]{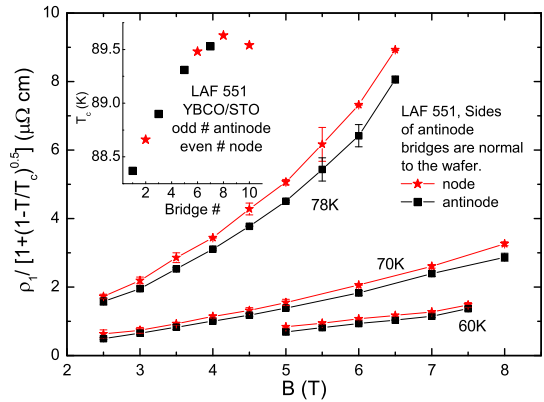}
\caption{(Color online) Averaged flux flow resistivity
$\rho_1/(1+\sqrt{1-T/T_c})$ versus magnetic field B on LAF551 at
60, 70 and 78 K, for the node (stars) and antinode (squares)
directions. The inset shows the distribution of the $T_c$ values
on the wafer. }\label{fig2}
\end{figure}
\begin{figure}
\includegraphics[width=8cm]{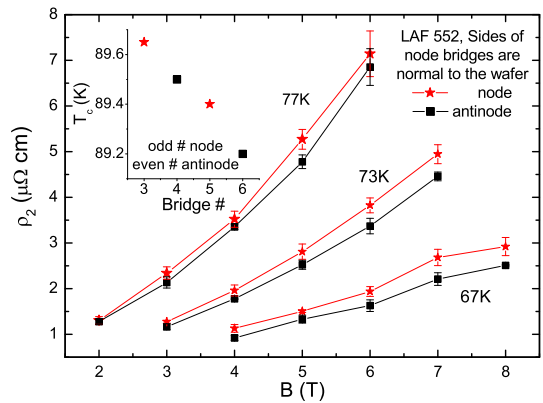}
\caption{(Color online) Averaged flux flow resistivity $\rho_2$
versus magnetic field B on LAF552 at 67, 73 and 77 K, for the node
(stars) and antinode (squares) directions. The inset shows the
distribution of $T_c$ for the four measured bridges on the wafer.
}\label{fig3}
\end{figure}
\indent Plots of the flux flow resistivities
$\rho_1/(1+\sqrt{1-T/T_c})$ and $\rho_2$ (calculated from $R_1$
and $R_2$) versus the magnetic field B are shown in Figs. 2 and 3
for the two wafers, respectively, at various temperatures. The
spread of the mid-point $T_c$ (where the low bias $dR/dT$ is
highest) of the different bridges on the two wafers are given in
the insets to these two Figures. The missing data for bridges \#4
and \#9 in Fig. 2 is due to bad contacts. The relatively large
spread of the $T_c$ values on this wafer of about 1.2 K, and the
asymmetry relative to the center of the wafers is due to an
off-center alignment of the laser ablated plume on the wafer.
Nevertheless, bridges \#5-8 on LAF551 (Fig. 2) and bridges \#3-6
on LAF552 (inset of Fig. 3, in a reversed order of the bridges)
have a much smaller $T_c$ spread of about 0.4K. Thus the FFR in
these bridges was measured and compared, and no systematic
correlation with the corresponding $T_c$ values was found. In
Figs. 2 and 3 average values are shown of the two measured node
bridges and the two measured antinode bridges on each wafer. The
errors in these figures are due mainly to this averaging
procedure. These figures show that $\rho_1/(1+\sqrt{1-T/T_c})$ and
$\rho_2$ increase almost linearly with B at low temperatures, but
more than linearly at higher temperatures. In all cases, a clear
anisotropy is observed where the $\rho_1/(1+\sqrt{1-T/T_c})$ and
$\rho_2$ values along the node direction are larger than along the
antinode direction. The fact that the \textit{linear} term in Eq.
(1) $R_1\propto \rho_1$ was found to be anisotropic seems to
contradict the four-fold symmetry of the twinned crystal. We note,
however, that $R_1$ (and also $R_2$ and $V^*$) are voltage
dependent (see inset (b) to Fig. 1). The results of this inset
were obtained by fitting the antinode I-V data using Eq. (1) near
the instability regime at various voltage segments of $V\pm 5$ mV.
The segments contained between 6 data points (above the maximum of
I and just below the voltage jump) to 17 points at the lowest
voltage bias shown. All fits had $R^2$ of better than 0.999 (solid
symbols). Also shown, are results of two inferior fits with 48 and
65 data points (empty symbols), and as one can see, the resulting
parameters still follow the general behavior. We thus conclude,
that $R_1$ is \textit{not} a linear term of the FFR, and therefore
does not have
to obey the symmetry of the heavily twinned crystal.\\
\begin{figure}
\includegraphics[width=8cm]{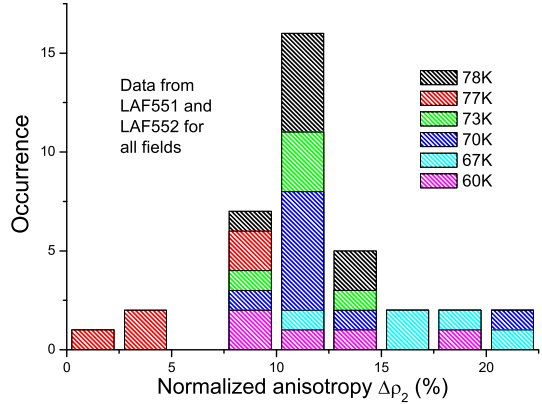}
\caption{(Color online) Histogram of the anisotropy results on
both wafers
$\bigtriangleup\rho_2$=$\rho_2$(node)-$\rho_2$(antinode)
normalized by the average value of $\rho_2$ for all fields and
temperatures. }\label{fig4}
\end{figure}
\indent Fig. 4 shows a histogram of the normalized anisotropy
$\bigtriangleup\rho_2$=$\rho_2$(node)-$\rho_2$(antinode) for all
the $\rho_2$ data on the two wafers. In this figure
$\bigtriangleup\rho_2$ is normalized by the average value
[$\rho_2$(node)+$\rho_2$(antinode)]/2 at each field and
temperature. We first plotted the normalized anisotropy
$\bigtriangleup\rho_2$ versus field for the six different
temperatures, and found that it is basically field independent to
within the noise of the measurements. No systematic temperature
dependence of the anisotropy was found at 60, 67 and 77K as can be
seen in the histogram of Fig. 4. This apparently is due to noise
in these measurements. At 70, 73 and 78K however, one can see that
the anisotropy values are concentrated in a narrow range of about
$11 \pm 2$\%. We thus conclude that the anisotropy is temperature
independent in the temperature range of our measurements. We note
that the anisotropy is expected to vanish at high temperatures due
to increasing thermalization of the quasiparticles. Above 80 K
however, we could not observe the LO instability jump in the I-V
curves, to observe this effect.\\
\begin{figure}
\includegraphics[width=8cm]{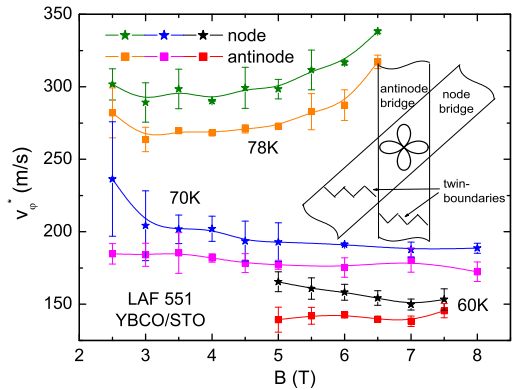}
\caption{(Color online) Critical vortex velocity $\rm v^*_\varphi$
calculated from $V^*$ of the fits to Eq. (1) versus B for the node
(stars) and antinode (squares) directions for the bridges of Fig.
2. The lines are guide to the eyes. The inset shows a schematic
drawing of the twin boundaries in the two kind of bridges.
 }\label{fig5}
\end{figure}
\indent Next, we turn to the critical voltage $V^*$ which
corresponds to the critical vortex velocity $\rm v^*_\varphi$
according to $V^*=|\textbf{v}^*_\varphi\times \textbf{B}|L$. In
Fig. 5 we plot average values of the critical vortex velocity $\rm
v^*_\varphi$ of the node (\#6,8) and antinode (\#5,7) bridges on
LAF551 versus magnetic field. In the node direction $\rm
v^*_\varphi$ is systematically larger (by about 10-20\%) than in
the antinode direction. In LAF552 this anisotropy is similar (not
shown). The $\rm v^*_\varphi$ values at each temperature show only
a weak field dependence, and are basically quite constant as was
found previously \cite{Dottinger}. Our conclusion therefore is
that the anisotropy in $\rm v^*_\varphi$ is consistent with the
anisotropy in $\rho_1$ and $\rho_2$, as all
have comparable percentage values.\\
\indent In the absence of a theoretical prediction for the
anisotropies of the flux flow resistivities $\Delta \rho_1$ and
$\Delta \rho_2$, and the vortex velocity $\rm \Delta v^*_\varphi$,
we shall present simple qualitative arguments for explaining them.
In principle, in the absence of pinning, the vortex velocity $\rm
v^*_\varphi$ is affected by the ability of the vortices to
transfer momentum to the quasiparticles. The Magnus or Lorentz
force that acts on a vortex constantly transfers momentum to it.
After a short acceleration, the vortex reaches a terminal velocity
$\rm v^*_\varphi$  in the viscous media while the extra momentum
has to be constantly dissipated to the quasiparticle excitations.
Thus the easier it is for a vortex to transfer momentum when it
moves in a certain direction, the faster it will move in this
direction. It was predicted theoretically that in a d-wave
superconductor the low energy excitations are located in the
vicinity of the nodes \cite{Volovik}. This was verified in
photoemission experiments where four small regions of
quasiparticle excitations were found on the Fermi surface in the
directions of the nodes \cite{Norman}. Our experimental
observations indicate larger flux flow velocity and FFR in the
node direction. Therefore, it seems that the momentum transfer by
vortices to the quasiparticles is more effective in the node
direction as compared to that along the antinode direction where
less quasiparticles are available for the momentum transfer
process. Clearly, a comprehensive theoretical analysis of this
issue is needed in order to make a detailed quantitative
comparison with the present results.\\
\indent Finally, we rule out the possibility that twinning is
responsible for the observed anisotropy. Twins in thin YBCO films
form a dense mosaic of elongated crystallites with boundaries
along the (110) and (1$\overline{1}$0) node directions as shown
schematically in the inset to Fig. 5. If the twin boundaries serve
as easy channels for flux flow, then flow in the anti-node bridge
is unhindered since there is always a Lorentz force component in
the direction of the motion. In the node bridge however, the flow
is blocked each time it reaches a boundary normal to its
direction. This should lead to a larger flux flow velocity and FFR
in the antinode direction as compared to the node direction. Since
this is opposite to the observed results, twins could not be the
source of our observations.\\
\indent In summary, we observed a clear anisotropy in the
nonlinear flux flow resistivities $\rho_1$ and $\rho_2$ of YBCO
which we attribute to the intrinsic anisotropy of the low energy
quasiparticle distribution on the Fermi surface of a d-wave
superconductor. The observed anisotropy in the critical vortex
velocity  $\rm v^*_\varphi$ is consistent with the anisotropy of
the measured flux-flow resistances.\\
\indent This research was supported in part by the German-Israel
Foundation, the Heinrich Hertz Minerva Center for High Temperature
Superconductivity, and the Israel Science Foundation, grants Nos.
1565/04 and 8003/02. The authors are grateful to A. Auerbach, Th.
Dahm, B. Ya. Shapiro and N. Schopohl for useful discussions.\\

\end{document}